# Holographic shear rheology of viscoelastic fluids

Siddhartha Gupta and Siva A. Vanapalli *

Dept. of Chemical Engineering,

Texas Tech University, Lubbock, TX

*Email: siva.vanapalli@ttu.edu

**Abstract**

In this study, we report the use of digital holography microscopy (DHM) for 3D-resolved flow kinematics and shear rheometry of viscoelastic polymeric fluids. We computationally reconstruct the recorded holograms to visualize the tracer imbued flow volume in microchannels, followed by implementation of particle tracking velocimetry (PTV) to quantitate spatially-resolved velocity fields in 3D. In order to select optimal parameters for DHM-PTV characterization of complex fluids, we studied the effect of hologram recording distance, seeding density and particle size. Using the optimal parameters, we show quantitative characterization of the shear rheology from the velocity fields without any a-priori assumptions of wall boundary condition or constitutive equation. The viscosity versus shear rate data for Newtonian and polyethylene oxide solutions could be measured in the range of ~ 0.05 - 20,000 $s^{-1}$ with just four input flow rates. This data from holographic shear rheometry was found to be in good agreement with computational fluid dynamics simulations and macrorheometry. The holographic shear rheology technique remained unaffected by wall-slip events and instead provided an avenue to quantitate slip severity. Finally, we discuss holographic visualization of particle migration in microfluidic flows which can limit flow field access while at the same time provide a fingerprint of the suspending fluid rheology.

## I. Introduction

The rheology of complex fluids is critical to determining the performance of consumer products such as paints and foods [1, 2]. In biological complex fluids as well, rheological properties play a crucial role in determining physiological outcomes [3-5]. The richness of complex fluids often leads to unique phenomena such as extensional thickening [6], shear banding [7] and elastic turbulence [8, 9]. Thus, high fidelity spatiotemporal flow field measurements can be quite useful to elucidate the physics of complex fluids and interpretation of rheological data.

Experimental methods to characterize flow kinematics and rheological properties of complex fluids often require integration of velocimetry techniques into rheometers. In this regard, studies have integrated scattering techniques [10], ultrasound [11], nuclear magnetic resonance [12], particle imaging velocimetry [13] and confocal imaging [14] into standard rheometers to obtain spatially resolved velocity fields. Such integration can be technically demanding due to the bulky rheometry setups and moreover acquisition of velocity data is often limited to two spatial dimensions.

Over the last decade there have been significant advances to miniaturize rheometry with the advent of microfluidic devices [15-17]. Several studies have shown shear [18-20] and extensional [21-23] rheology of viscoelastic fluids by using microfluidic devices with benefits of small sample volumes and access to high rates of fluid deformation. Many of these microfluidic rheometers quantitate rheology by measuring the relationship between pressure drop and flow rate relation [20, 24, 25]. Although such approaches enable determination of material properties, they do not provide information on flow structure or wall-slip.

Application of techniques to characterize flow kinematics in microfluidic devices can open new avenues for quantitating rheology [16, 26, 27] as well as for determining spatial structure of the flow [28-31]. Ideally, these techniques should be capable of accessing flow information in 3D, *i.e.* the three spatial dimensions due to the rectangular cross-section of microfluidic geometries, as well as with fast temporal resolution. Particle imaging velocimetry and confocal microscopy provide access to 3D resolved velocity fields but require mechanical scanning through the flow volume. As a result, they are limited by temporal resolution and best suited for steady flows.

For characterizing kinematics in 3D, digital holography microscopy (DHM) is well suited since it is a volumetric imaging technique allowing fast temporal resolution [32]. Holograms are reconstructed and computational scanning is performed to localize seeded particles in 3D [33]. Particle-tracking velocimetry (PTV) is then used to obtain 3D resolved velocity fields. This DHM-PTV has been previously used for micromixer flows [34], dean flows [35], flows on patterned surfaces [36], colloidal dynamics [26, 37], microchannel flows [38-40] and turbulence [41, 42].

Most prior studies of DHM-PTV have focused on Newtonian flows and its application to viscoelastic flows in microfluidic geometries is emerging. Shear banding and flow fluctuations due to worm-like micellar fluids in rectilinear microchannels have been studied using DHM-PTV [43]. In addition, viscoelastic flow around a confined cylinder has been mapped using DHM-PTV [44]. More recently, we have shown that DHM-PTV can be used to characterize 3D velocity fields in a hyperbolic contraction-expansion geometry [45].

In this study, we apply DHM-PTV to flow of viscoelastic polymeric fluids in linear microchannels and show that shear rheology can be directly obtained from the measured 3D velocity fields and imposed driving pressure. This approach referred to as Holographic Shear Rheology (HSR) not only measures nonlinear rheology of fluids but also informs on the presence of wall-slip and provides insights into viscoelastic particle migration. The shear viscosity curves from HSR are found to be in quantitative agreement with macrorheometry. Thus, HSR measures shear rheology of viscoelastic fluids without explicitly obtaining pressure drop and flow rate relation, but also provides quantitative information on wall-slip and flow structure.

**II. Working Principle of Holographic Shear Rheology**

To characterize the shear rheology of viscoelastic fluids, we employ holography-based particle-tracking velocimetry. The basic idea is to impose a known pressure drop in a thin microchannel and obtain velocity profiles using DHM-PTV. This enables calculation of shear stress versus shear rate relation, from which viscosity curves can be generated. In this section, we discuss this approach that forms the basis of our holographic shear rheology (HSR). First, we present the governing equations for quantitating shear rheology from velocimetry data. Second, we describe the details of implementation of the DHM-PTV analysis pipeline to obtain velocity fields.

## A. Quantification of shear rheology from velocimetry data

To determine shear rheology, we consider viscoelastic flow through a linear microchannel of length $L_{ch}$, height $h$ and width $w$. Our analysis follows that of ref. [46]. Ignoring external body forces and applying the Cauchy momentum equation gives $\frac{Du}{Dt} = \frac{1}{\rho}\nabla.\sigma$ where $\frac{D}{Dt}$ is the material derivative, $u$ is the local fluid velocity, $\rho$ is the fluid density, and $\nabla.\sigma$ is the divergence of the stress tensor. When the channel aspect ratio is small such that $h/w \ll 1$, and for steady unidirectional flow, the Cauchy momentum equation simplifies to $\frac{\partial \sigma_{xz}}{\partial z} = \frac{\partial P}{\partial x} = \frac{\Delta P}{L_{ch}}$ where $x$ and $z$ are the stream-wise and depth-wise coordinates respectively, and $\frac{\partial P}{\partial x}$ is the stream-wise pressure gradient that can be determined from the known imposed pressure drop $\Delta P$. Upon integration, the local shear stress becomes: $\sigma_{xz} = \left(\frac{\Delta P}{L_{ch}}\right)(z - z_{\dot{\gamma}=0})$ where $z_{\dot{\gamma}=0}$ indicates the plane of zero shear stress (or maximum velocity) which may not be at the mid-plane. The local shear rate can be determined from the measured depth-wise velocity profile as $\dot{\gamma}_z = \frac{\partial u_x}{\partial z}$. Thus, knowing the shear stress versus shear rate data, shear viscosity curves can be determined since shear viscosity $\mu = \frac{\sigma_{xz}}{\dot{\gamma}_z}$.

The above analysis has parallels to that of slit or capillary rheometry [47] where pressure drop versus flow rate relations are used to quantitate shear rheology. Microfluidic viscometers also use such relations to characterize shear rheology [16, 20, 48]. In HSR, we do not measure volumetric flow rate but instead calculate local velocity gradients from the velocity profile. As a result, in situations where wall-slip is present, viscometers that rely on measuring volumetric flow rate can be prone to error. However, wall-slip does not affect the HSR approach since local shear rate is obtained rather than calculating mean shear rate from flow rate. The importance of wall-slip in viscoelastic microflows is further discussed in Sec IV D.

## B. Implementation of the DHM-PTV analysis pipeline

To characterize velocity profiles of viscoelastic fluids, we developed a DHM-PTV analysis pipeline that consists of the following steps (i) the fluid is seeded with non-deformable microparticles (ii) inline holography records the tracer imbued volume as 2D holograms (iii) the

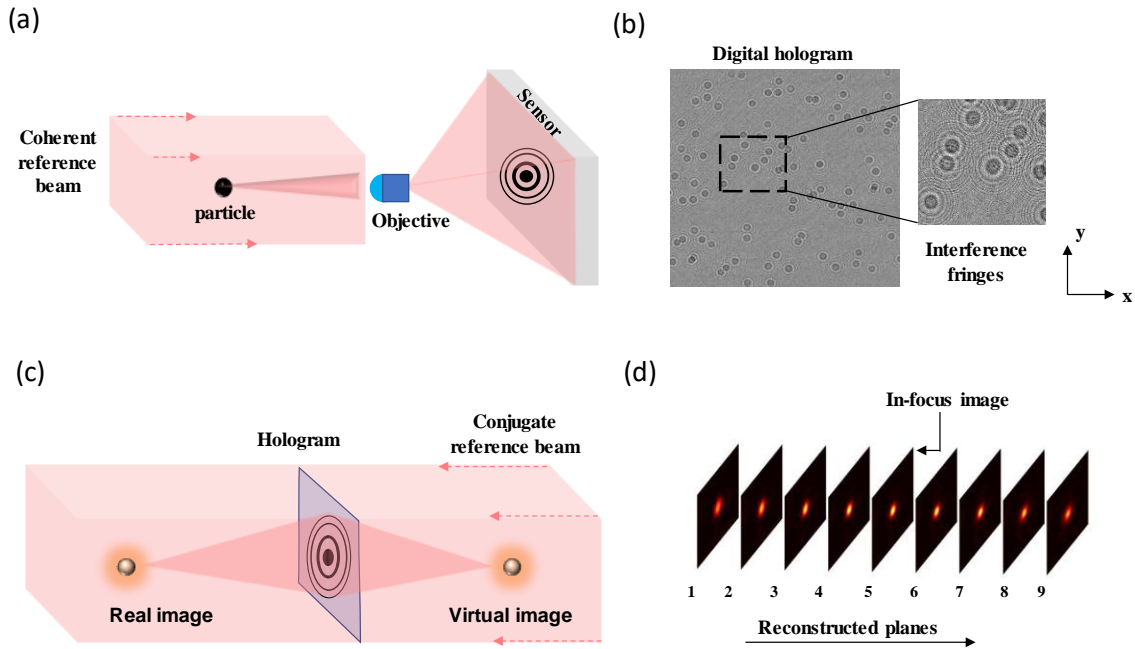

**Figure 1.** Principle of Digital Holography Microscopy (DHM) (a) Digital hologram recording on a digital sensor with collimated laser beam. The interference between the reference beam and forward scattering from the particle leads to interference fringe patterns. The fringes are magnified by a microscope objective prior to recording on the sensor. (b) Cleaned hologram of dilute particle (diameter = 2 µm) suspension flowing in a PDMS micro-slit recorded at 20X magnification and at 512 µm x 512 µm field of view. Inset shows zoomed-in fringe patterns. (c) Principle of digital reconstruction of object field from the recorded hologram. The conjugate of reference beam is numerically imposed on the recording and convolution with a transfer function yields reconstruction of the 3D volume. (d) Plane wise reconstruction of the intensity of a particle in the image volume.

scattering field of individual particles in the flowing volume is recovered by digital reconstruction (iv) particle centroid locations in 3D are identified in the image volume (v) trajectories are linked frame-wise using PTV [49] and the velocity field is determined. A detailed explanation of these different steps is provided below:

**In-line recording of digital holograms.** In-line digital hologram recording is done by illuminating sample space with a coherent reference beam of collimated laser light and recording the forward interference patterns on a sensor located perpendicular to the reference beam (Fig. 1a). The forward scattering from the object *i.e.* object wave and the reference wave interfere in the focal plane of the microscope objective located beyond the sample volume and gets recorded as fringe patterns (Fig. 1b). Holograms are magnified by a microscope objective prior to recording to enhance fringe resolution and improve the depth wise accuracy during reconstruction. The intensity distribution of the raw hologram is denoted as $I_{i,raw}(x_h, y_h)$ where $x_h, y_h$ denote pixel coordinates on the 2D image and $i$ is the index corresponding to the hologram number in the recorded video.

**Digital reconstruction of the particle scattering field.** The raw holograms are digitally reconstructed by computationally imposing a conjugate reference beam and calculating the forward scattering (Fig. 1c). This process effectively provides 3D visualization of the flow volume with particles appearing as bright scattering regions against the external background. Operationally, there are two steps to the digital reconstruction process. First, individual raw holograms are cleaned by removing noise using a background hologram that is obtained by averaging a sequence of typically 100 holograms. Mathematically, $I_{i,clean} = I_{i,raw} - I_{bgrd}$ where the intensity distribution of the background hologram $I_{bgrd} = \frac{1}{N}\sum_{i=1}^{N} I_{i,raw}$. Second, reconstruction is done on the cleaned holograms using the angular spectrum method as it does not have a minimum distance requirement [34, 50] and allows computationally efficient reconstruction with improved signal to noise ratio [51, 52]. Under the angular spectrum method, the field propagation is expressed as a linear filtering of the angular spectrum of the original field. The reconstructed complex amplitude $A_i(x, y, z)$ is obtained by convolving the cleaned hologram $I_{i,clean}(x_h, y_h)$ with the free space transfer function $h_z(x, y, z; x_h, y_h)$ [50, 53-55], i.e.

$$A_i(x, y, z) = I_{i,clean}(x_h, y_h) \otimes h_z(x, y, z; x_h, y_h) \tag{1}$$

Here, $x$ and $y$ denote the spatial coordinates in the reconstructed plane (which are also the same as the spatial coordinates in the flow) and $z$ indicates the depth-wise position of the reconstruction plane. The convolution is implemented using Fast Fourier Transform (FFT) based calculations as:

$$A_i(x, y, z) = \mathcal{F}^{-1}[\mathcal{F}\left(I_{i,clean}(x_h, y_h)\right) \times \mathcal{F}(h_z(x, y, z; x_h, y_h))] \tag{2}$$

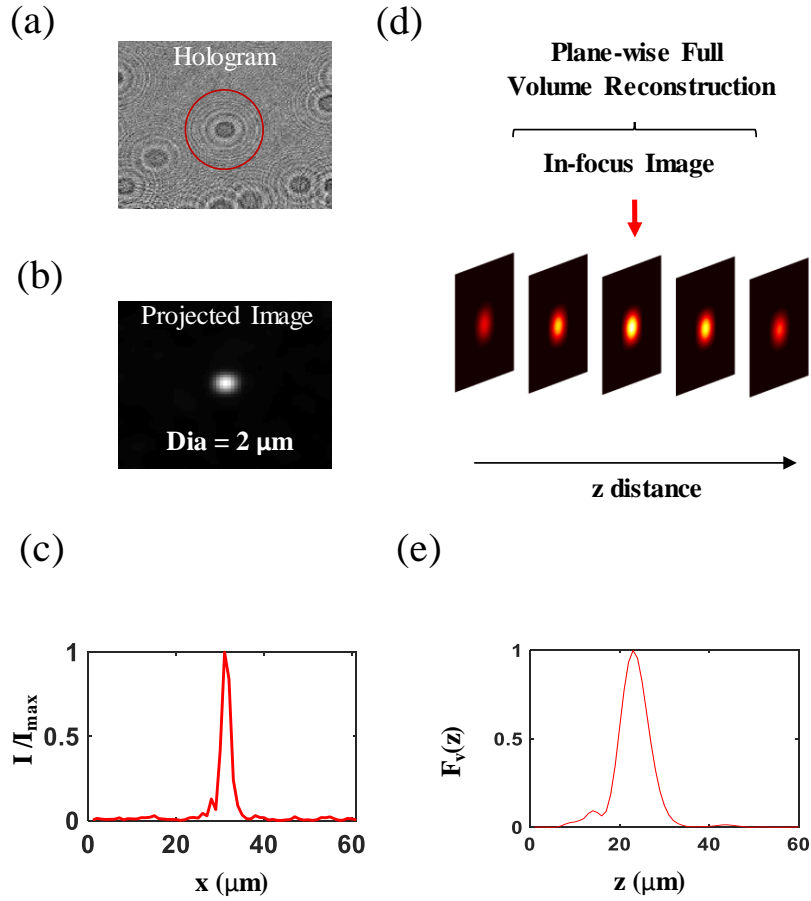

**Figure 2.** Locating the particles in three dimensions. (a) A particle is identified in a cleaned hologram. (b) The reconstructed planes are projected in a 2D image that is used to segment the objects as regions of bright intensity. (c) The peak lateral intensity profile (I /Imax) is used to get the planar centroid location of the particle. (d) The full plane wise stack of the intensity volume is reconstructed. (e) The Laplacian of the axial intensity $F_v(z)$ is calculated along the depth, and its maxima is used to determine the z location of the particle centroid.

Here, $\mathcal{F}$ and $\mathcal{F}^{-1}$ denote the FFT and inverse FFT respectively. The intensity distribution corresponding to the 3D particle field is then calculated as $I_{i,volume}(x,y,z) = |A_i(x,y,z)|^2$, which is shown in Fig. 1d as a series of reconstructed planes.

**Particle localization in 3D space.** The reconstructed planes are used to locate the centroids ($x_c$, $y_c$, $z_c$) of the particles in 3D. To determine the ($x_c$, $y_c$) location of a particle in the raw hologram (Fig. 2a), the maximum intensity at every pixel coordinate is obtained by scanning all the planes and projecting onto a 2D image (Fig. 2b). The peak in the intensity profile of the projected image (Fig. 2c) is identified as ($x_c$, $y_c$,). To identify the z-coordinate of the centroid, the plane of best focus is determined. This is done by performing a Laplacian operation on the intensity distribution in the image volume (Fig. 2d), i.e. $F_v(z) = \sum_{x,y} \left(\nabla^2 I_{i,volume}(x,y,z)\right)^2$, with the summation carried over a 3 x 3-pixel grid around the ($x_c$, $y_c$) location in each reconstruction plane. The plane of focus is chosen as the plane where $F_v(z)$ is maximized and the z-location of this focal plane is chosen as $z_c$ (Fig. 2e). Thus, the tracer particles in the flow volume are localized in 3D.

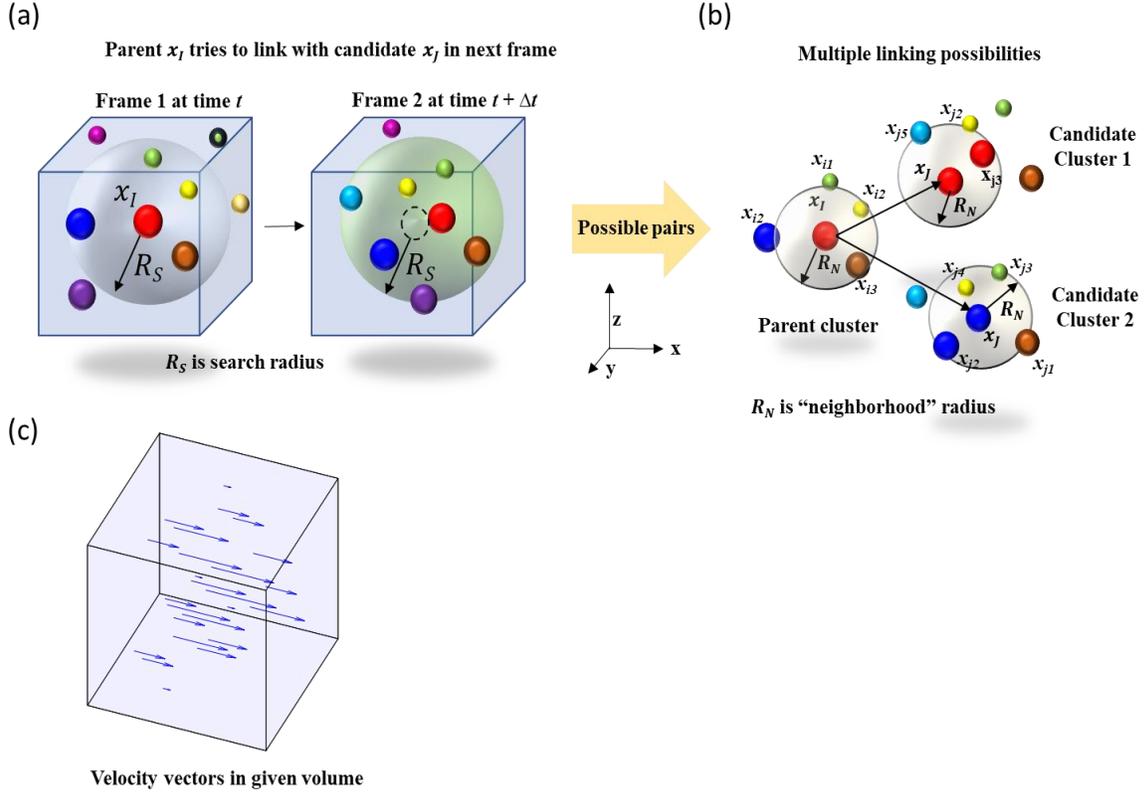

**Figure 3.** The particle tracking velocimetry approach is based on the method proposed by Ishikawa et al., 2000 [56]. (a) First a parent particle $x_I$ is chosen in frame 1 at time t and an attempt to link its trajectory is made by finding it in the next frame at time t + Δt as candidate $x_J$ within a search radius R$_s$. (b) Multiple linking possibilities arise in cases of multiple candidate particles in the search radius Rs. A cluster of search radius R$_n$ is identified around the parent and its candidates in the next frame and all the participants of the cluster pair are linked piecewise to obtain the velocity gradient tensor matrix for each combination. The most probable link is estimated using a least squares minimization strategy. (c) The trajectories from the pairings are stored as coordinates of the tail and the head of the displacement vector.

**Particle tracking velocimetry to obtain velocity field.** Once particle centroids are established, we map the tracer displacement field by evaluating trajectories with PTV. The PTV algorithm used in this study is based on calculating velocity gradient tensors (VGT) as proposed by Ishikawa *et al.* which was chosen for its suitability to map 3D flows [49, 56]. The basic idea entails estimating the flow feature for a parent particle having a local neighborhood by correlating it with a possible candidate in the next frame having a similar neighborhood and calculating the velocity gradient tensor matrix between the two particles in the respective frames. Briefly, a parent particle $x_I$ is chosen in the first frame and potential candidates $x_J$ for linking trajectories are found in the next frame within a search radius $R_s$ (Fig. 3a). Next, a cluster of neighborhood radius $R_n$ is formed around the first frame particle having neighbors $x_{in}$, where n denotes the index of the neighbor, and similar clusters are assumed around the candidates $x_J$ in the second frame (Fig. 3b) having neighbors $x_{jn}$. The choice of $R_n$ is made to retain at least 2-3 particles in each cluster and every cluster pair is evaluated by calculating the VGT tensor via a least square's minimization approach. The minimization exercise is formulated as:

$$E_{IJ} = \sum_{k=1}^{n} |X_{J,K} - AX_{IK}|^2 \qquad (3)$$

Here, the matrix $A = I + \partial u(x_I)\Delta t$ includes the velocity gradient tensor $\partial u(x_I)$ and the unit matrix $I$ and $X_{J,K}, X_{I,K}$ are distances of cluster centers from their neighbors. The best possible pairing is decided by minimal value of $E_{IJ}$. The pairings are stored as tail and head coordinates of the trajectory displacement vector (Fig. 3c).

**Postprocessing of velocity vector data.** The DHM-PTV output is susceptible to statistical noise intrinsic to the linking process, fringe distortion at walls and noisy reconstruction. Additionally, vector generation is sensitive to particle distribution within the flow which can be sparse. Therefore, we construct a regular grid aligned with the flow cross-section to project the PTV velocity vector data. The projection is consistent with steady state flow invariance in the streamwise direction. The PTV data is median and gaussian filtered to remove outliers before interpolation onto the uniform grid for calculation of the flow field. Each velocity vector within a grid element is ascribed to the center and only those grid cells are considered which contain at least 8-12 velocity vectors inside the grid element.

### III. Experimental Methods

**A. Digital holography microscopy**

The in-line digital holography microscopy set up used in this study including the optical train and imaging system is the same as that of our previous works [57, 58]. The holograms of the microchannel flow are recorded in the focal plane of a 20X magnification microscope objective (NA = 0.45, Olympus). The magnified fringe patterns from the hologram are captured on a CMOS camera (Phantomv310, Vision Research) with a field-of-view (FOV) of 512 x 512 pixels. This imaging system yields a resolution of 1 µm per pixel. An exposure time of 9 – 11 µs was employed, and the frame rate (24 – 11001 fps) was controlled so that the fastest particle traversed 5 – 40 pixels between frames. The recording distance, *i.e.* distance between the microchannel floor and the focal plane of the microscope objective was maintained at 100 µm unless otherwise stated.

Holographic reconstruction was done with an inter-plane spacing of 1 µm. While accurate velocity fields were obtained from ~ 1000 holograms, we recorded and analyzed ~ 10,000 holograms to improve PTV statistics especially when sparse particle fields can be present in the channel domain due to particle migration in viscoelastic flows, as further discussed in Sec. IV D. The hologram processing and PTV analysis were performed using custom routines with parallel computing capability written in MATLAB (MathWorks). Computational processing was done on a desktop (XPS 8930, Dell Inc.) running Windows OS on multiple cores (Intel Core i7-8700K CPU @ 3.70GHz, 3696 MHz, 6 Cores). Each hologram pair required ~ 2 sec processing time [this includes reconstruction and PTV analysis. A 10,000 hologram video required 7 hours of computational time to yield velocity vector data.

**B. Sample preparation**

The choice of Newtonian test fluid was DI water. For polymeric fluids, polyethylene oxide (PEO) of reported molar mass of 4 x $10^6$ g/mol (WSR301, DOW) was used which had an overlap concentration c* ~ 620 ppm [59]. A semi-dilute stock polymer solution of 1 wt% was prepared by dissolving PEO in DI water and stirring at 85 rpm for 48 hours using a magnetic stir bar. The stock solution was stored at 4 °C wrapped in an aluminum foil to prevent photo-degradation. The stock was thence diluted serially to 0.5, 0.25 and 0.025 wt.% prior to experiments. Our optimal seeding density for the test fluids was ~ 0.14 v/v.% (or 9 x $10^6$ particles/mL) for polystyrene microspheres of diameter 2 µm (density 1.05 g/cm$^3$, PS 19814-15, Polysciences). This resulted in particle number density of ~ 100 particles in the FOV for the thin-slit microchannel and ~ 50 particles for the microchannel with the square cross-section. To evaluate the effect of particle size on DHM-PTV performance, we also tested particles of 3 µm (PS 17134-15, Polysciences) and 6 µm (PS 07312, Polysciences) diameter.

**C. Rheological characterization**

The shear viscosity curves for all the PEO solutions were measured on a macro-rheometer (AR2000, TA instruments) using the double gap geometry at the temperature (21 – 23 ºC) of the microfluidic flow experiments. In addition, their relaxation time was determined using dripping-on-substrate rheometry [60] and our set up for this measurement is identical to that reported recently [45]. The measured relaxation times for 1, 0.5, 0.25 and 0.025 wt.% PEO solutions were 240.1 ± 20.5, 182 ± 18, 55 ± 3.9 and 7.1 ± 0.15 ms respectively.

**D. Microfluidic device fabrication**

The experiments were done in linear microchannels with thin-slit (rectangular) and square cross-sections. To fabricate the microfluidic channels, we used SU8-based soft lithography [61]. Negative photomasks designed in AutoCAD were printed. Next, an SU-8 mold was made using soft lithography on a 3'' silicon wafer. The height of the channels was controlled during the spin coating process and was targeted to be 50 µm for the 500 µm wide thin-slit and 100 µm for the square channel. Polydimethylsiloxane (PDMS) was prepared by mixing cross-linker and base (Sylgard-184 Silicone Elastomer kit, DOW) in a 1:10 wt.% ratio and degassed before being poured on the SU-8 mold. The mold was cured in an oven for 5-6 hours at $65^0$C after which the PDMS chips were peeled off and characterized using a microscope (CKX41, Olympus) to determine the height variation. Post characterization, the height of the thin-slit microchannel was found to be ~ 44 µm and width 500 µm, whereas, for the square microchannel the height was found to be 103 µm and the width to be 105 µm. The error in these spatial dimensions corresponds to the optical resolution of ± 1 µm. Inlet and outlet reservoirs were defined by punching holes and the channels were irreversibly bonded to a glass slide (25mm × 75mm × 1mm, Fisher) after exposing the bonding surfaces of the PDMS device and glass slide to plasma (Harrick Plasma) for 2 minutes.

### E. Flow experiments

For HSR experiments, constant pressure at the inlet of the microfluidic devices was imposed using a pressure controller (MFC8-FLEX4C, Fluigent Inc.). The pressure drop across the device was varied discretely from 0 - 355 mbar. For validating the flow kinematics, constant volumetric flow rate of 500 µL/hr was imposed in the thin-slit and square microchannel devices using syringe pumps (PHD2000, Harvard Apparatus). After starting the flow, a stabilization time ~ 2 - 10 minutes was allowed. DHM imaging was performed at ~ 115$D_H$ and ~ 100$D_H$ for the thin-slit and square microchannel geometries respectively, which is sufficient for flow stabilization at the Reynold's numbers of our experiments [62]. Here, $D_H$ is the hydraulic diameter of the microchannel.

For the flow conditions used in the study, the Reynolds number varied from $Re$ ~ $10^{-5}$ – 32. Due to the shear-thinning nature of the PEO solutions, we defined $Re = \frac{\rho \bar{v}^{2-n} D_h^n}{K\left(\frac{3n+1}{4n}\right)^n 8^{n-1}}$ where $\bar{v}$ is the average flow velocity, K (Pa.s$^n$) is the power law prefactor and n is the power law exponent [63]. In our study, the Weissenberg number varied from $Wi$ ~ 1 – 161, with $Wi = \lambda \dot{\gamma}_c$ where $\dot{\gamma}_c$ (= $\frac{2\bar{v}}{h}$) is the characteristic shear rate and $\lambda$ is is the relaxation time of the fluid. Finally, the Elasticity number defined as $El = \frac{Wi}{Re}$ varied from ~ 3.5 - $10^4$.

### F. Computational fluid dynamics

To validate velocity profiles obtained from DHM-PTV, we performed finite volume-based simulations using the computational fluid dynamics (CFD) package Fluent (Ansys). The CFD simulations were validated against analytical expressions for Newtonian fluid in both the thin-slit and square microchannels to optimize meshing and model setup. The power-law fluid model was used to simulate viscoelastic flow in the microchannels, and the resulting velocity profiles were compared with those obtained from DHM-PTV.

## IV. Results and Discussion

### A. Optimization of system parameters for DHM-PTV

Successful determination of the 3D velocity profiles requires optimization of system parameters which might include those from the holography setup as well as those pertaining to PTV analysis. Here, we considered experimental optimization of the following system parameters: the recording distance for the holograms, particle size and particle concentration. The optimization was pursued by keeping one parameter fixed and varying the other two and evaluating the degree of error $\Delta V_{rms}$ between the measured and theoretical velocity profile for a Newtonian fluid in a microchannel (see Eqn. 4, where $\dot{Q}$ is the flow rate). Here, $\Delta V_{rms}$ represents the root mean-squared (RMS) error of the mid-plane width-wise velocity profile calculated from the measured values and the analytical result for flow in a rectangular channel [64]. The width-wise profile was chosen for calculating RMS error since it has more measured values enabling better statistical comparison.

$$u_x(y,z) = \frac{\frac{48\,\dot{Q}}{\pi^3 hw}\sum_{n,odd}^{\infty}\frac{1}{n^3}\left[1-\frac{\cosh\left(n\pi\frac{y}{h}\right)}{\cosh\left(n\pi\frac{w}{2h}\right)}\right]\sin\left(n\pi\frac{z}{h}\right)}{\left[1-\sum_{n,odd}^{\infty}\frac{192h}{n^5\pi^5 w}\tanh\left(n\pi\frac{w}{2h}\right)\right]} \quad (4)$$

In this study, we used a thin-slit microchannel (Fig. 4a) and optimized the system parameters. The tested conditions shown in Fig. 4b-d include: recording distance $Z_{rec}$ = 100, 600, 1200 μm; particle size $D_p \approx$ 2, 3 and 6 μm; and particles per frame (or image volume) $N_o \approx$ 50, 100, 300. Below, we elaborate on the results from this optimization study.

**Hologram recording distance**. In a holography system, despite the large depth of field afforded in comparison to conventional microscopy, the recording distance $Z_{rec}$ needs to be optimized because small recording distances are susceptible to noise from twin image formation [65] whereas large separation between the object and hologram recording plane position suffers from aberrations led by finite numerical aperture (NA) objective. Moreover, in case of microfluidic channel flows, light rays refract from the flow media as well as the glass substrate and subtend a reduced light cone on the hologram plane (FOV) as the distance from the object is increased, reducing the effective NA of the system [34]. Thus, there is a need to select the optimal recording distance.

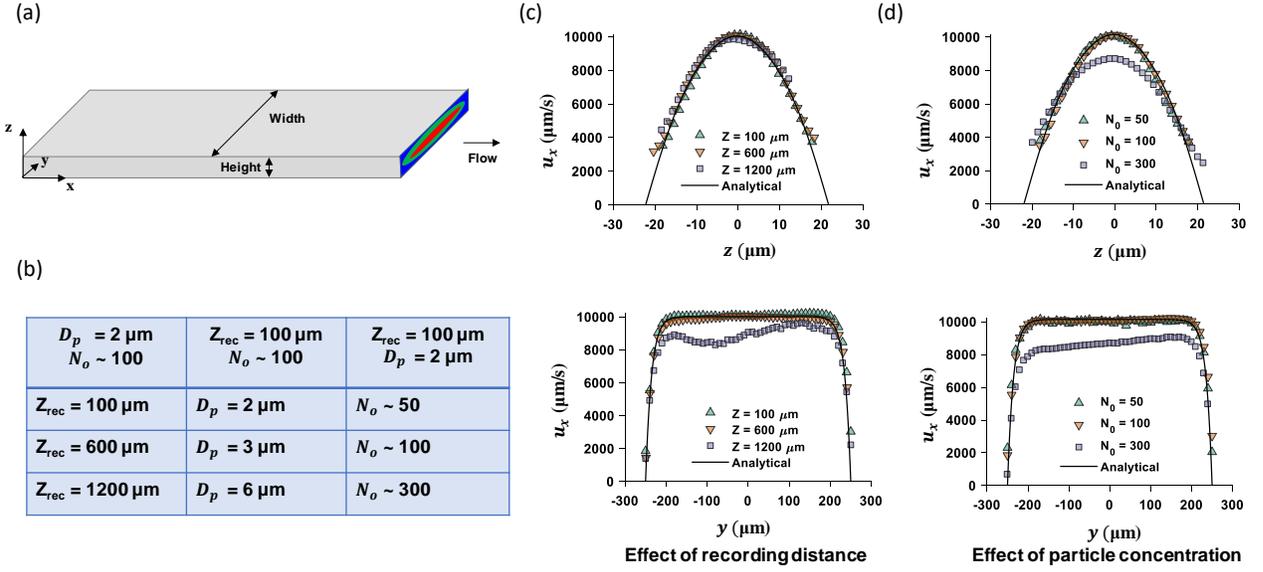

**Figure 4.** Optimization of DHM-PTV in the thin slit. (a) Thin-slit microchannel geometry used for the optimization studies. (b) Three sets of trials were conducted for optimizing the DHM-PTV analysis which include particle size $D_p$, particle number density $N_o$ and the recording distance $Z_{rec}$. The normalized RMS error in velocity estimation from DHM PTV versus the analytical result for different trials - (c) the effect of recording distance and (d) the effect of particle seeding density. In (c) and (d), the top and bottom panels are the depth-wise and width-wise mid-plane velocity profiles. Flow rate is fixed at 500 µL/hr corresponding to Re = 0.59.

In Fig. 4c, we show the measured mid-plane width-wise and depth-wise velocity profiles obtained for fixed particle size $D_p$ = 2 µm and particle per frame $N_o$ ~ 100, but varying $Z_{rec}$ = 100, 600 and 1200 µm. We find that the RMS error is ~ 2.5% of the maximum velocity ($V_{max}$) for $Z_{rec}$ = 100 and 600 µm, however, it increases to ~ 10% for $Z_{rec}$ = 1200 µm as reconstruction suffers from optical aberrations that arise due to the reduced angular range of light rays incident on the FOV as well as from using a finite NA objective.

**Particle size**. The particle size is important because, depending on the diameter $D_p$ the particle may lead or lag the flow [66, 67] and migrate towards centerline or the wall [68]. The finite size limits closest approach to the wall restricting information access from slower streamlines. We measured the velocity profiles for three different particle sizes $D_p \approx$ 2 µm, 3 µm and 6 µm, while maintaining $N_o$ ~ 100, and $Z_{rec}$ = 100 µm. We find that the RMS error remains under 2.5 % and does not vary significantly for the particle sizes considered (data not shown), although we observe that for $D_p \approx 6$ µm, only < 73% of the channel depth could be probed due to exclusion of slow streamlines and hydrodynamic resistance due to increased particle size (confinement) [66, 67].

**Particle seeding density**. An important factor for successful DHM-PTV analysis is the particle seeding density since it impacts the shadow density and vector yield. The shadow density $S_d$ refers to the degree to which scattering intensity or 'shadows' from particles located in the 3D image volume overlap when projected onto a 2D frame [69, 70]. A measure of shadow density is $S_d = \frac{N_o(D_p)^2}{wL_{fov}}$, where $w$ is the width of the channel, $L_{fov}$ (= 512 µm) is the stream-wise length of the field-of-view. It is clear that shadow density depends on the particles per frame $N_o$ as well as particle size $D_p$. Here, we tested the influence of seeding density by maintaining , $D_p = 2$ µm (and $Z_{rec} = 100$ µm), and varying $N_o$ ~ 50, 100 and 300. For these conditions $S_d$ ranges from 0.08% - 0.48%. In general, reconstruction efficiency decreases with increasing $N_0$ or $S_d$ [70]. In addition, this loss in reconstruction efficiency can lead to missing particles between frames hampering PTV vector yield. Alternatively, decreasing $N_o$ significantly reduces the vector yield necessitating more holograms and greater processing times. Thus, there is a need for optimizing particle seeding density. The velocimetry results from this optimization are shown in Fig. 4d. The velocity profiles deviate significantly from the analytical result for $N_o$ ~ 300 yielding an RMS error of 15%. In contrast, the RMS error for $N_o$ ~ 50 and ~ 100 is less than 2.5 %.

In summary, our optimization studies of system parameters revealed the conditions that yield RMS error of 2.5% or less. In this study, we chose $Z_{rec} = 100$ µm, $D_p = 2$ µm and $N_o = 100$ particles per frame as the optimal operating parameters for DHM-PTV implementation.

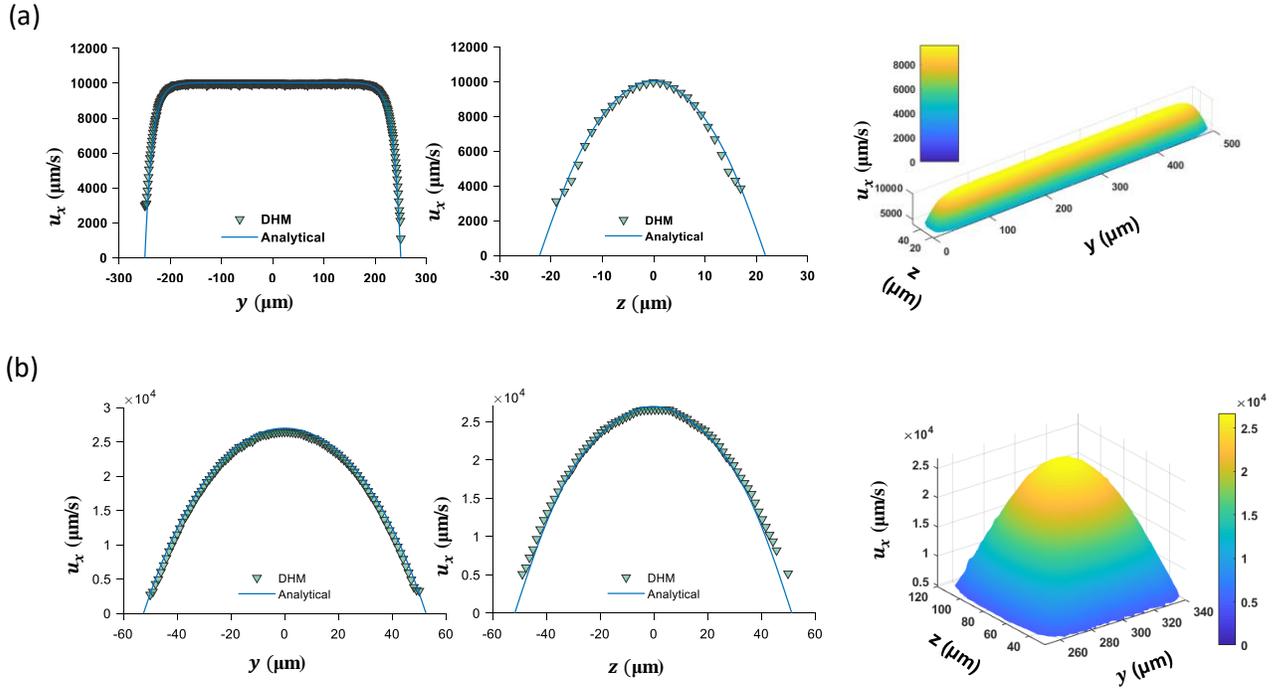

**Figure 5.** 3D velocity profile for Newtonian fluid (a) The results for midplane velocity profiles along the height and width of the thin-slit channel alongside 3D convex envelope of velocity over the flow cross section (Re = 0.59). (b) Corresponding results in case of a square microchannel (Re = 1.38). The color scale indicates velocity variation. The Newtonian fluid is water and volumetric flow rate is 500 µL/hr.

### B. Validation of DHM PTV for flow field characterization

To further validate that the optimal parameters have been identified for both Newtonian and viscoelastic fluids, we measured the 3D velocity profiles for water and 0.5 wt. % PEO and compared them with Eqn. 4 for Newtonian flows and CFD simulations for the case of viscoelastic flows. We performed these comparisons in both the thin slit and square microchannels by imposing a constant flow rate and measuring the velocity profiles.

As shown in Figure 5a,b, the mid-plane velocity profiles for water in the span-wise and depth-wise directions agree well with the analytical result from Eqn. 4 [64] for a thin slit as well as a square microchannel geometry. The envelopes of 3D velocity profiles are also shown and display a distinct wedge-like appearance in the thin-slit and a paraboloid for the square microchannel.

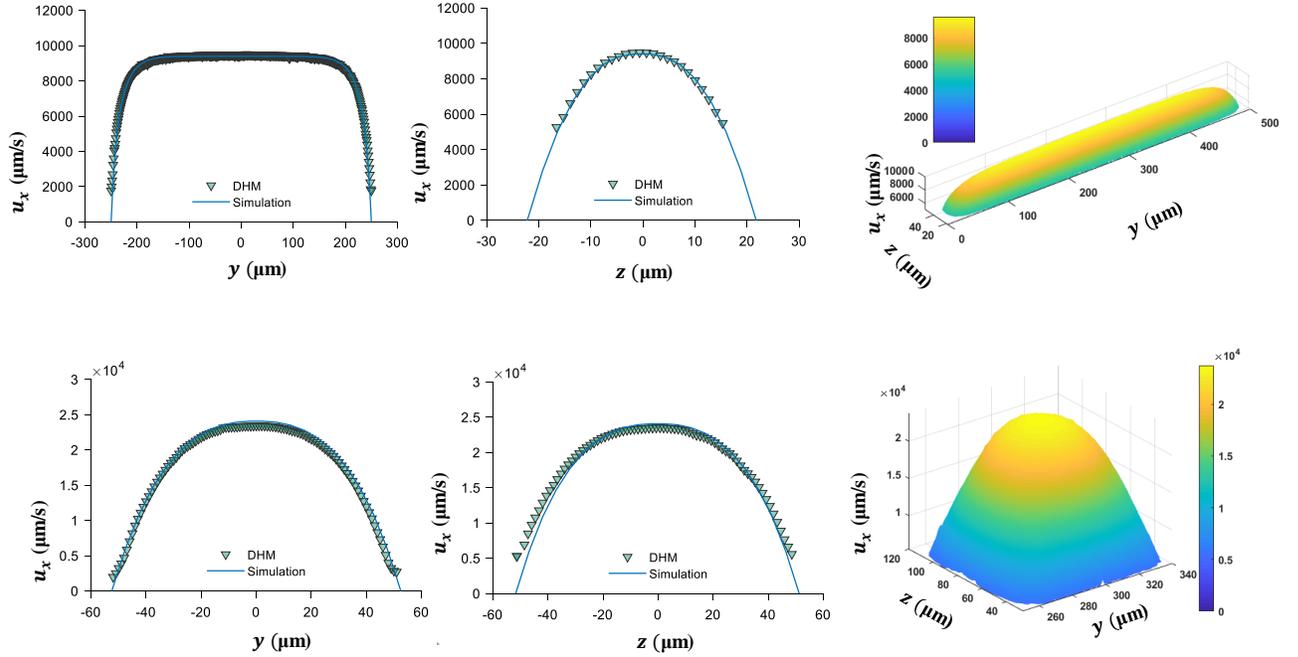

**Figure 6.** 3D velocity profile for viscoelastic 0.5 wt% PEO solution. (a) Mid-plane velocity profiles along the depth and width of the thin-slit channel alongside 3D convex envelope of velocity over the flow cross section (Re = 0.038, Wi = 56). (b) Corresponding results in case of a square microchannel (Re = 0.07 and Wi = 35). Color scale indicates velocity variation. The volumetric flow rate was fixed at 500 µL/hr.

Next, results for flow of viscoelastic solution of 0.5 wt.% PEO are shown in Fig. 6a for the thin-slit and for the square microchannel in Fig. 6b. Also, shown are the results from the CFD simulation using a power-law fluid model with $K$ and $n$ obtained from conventional rheometry. The experimental and simulation results are in good agreement. The viscoelastic flow exhibit a more blunted velocity front that departs from the parabolic Newtonian flow profiles due to shear thinning.

We note that the depth-wise velocity profiles truncate more so than the span-wise velocity profiles in both the geometries and for both the fluids. In particular, the depth-wise velocity profiles in the thin slit truncate markedly by ~ 3 - 5 µm for Newtonian flows and ~ 4 - 6 µm for viscoelastic flows from the channel roof and floor. The reason for this truncation is due to sparse vector fields near the wall. As discussed further in Sec. IV D, we find that for the conditions explored in this study, particles migrate away from the walls [71-75], creating near-wall fluid regions that are sparsely populated with particles, making it difficult to faithfully extract velocity vectors close to the wall. Despite this limitation, as we show in the next section, reliable shear viscosity curves can be obtained from the available flow kinematics data.

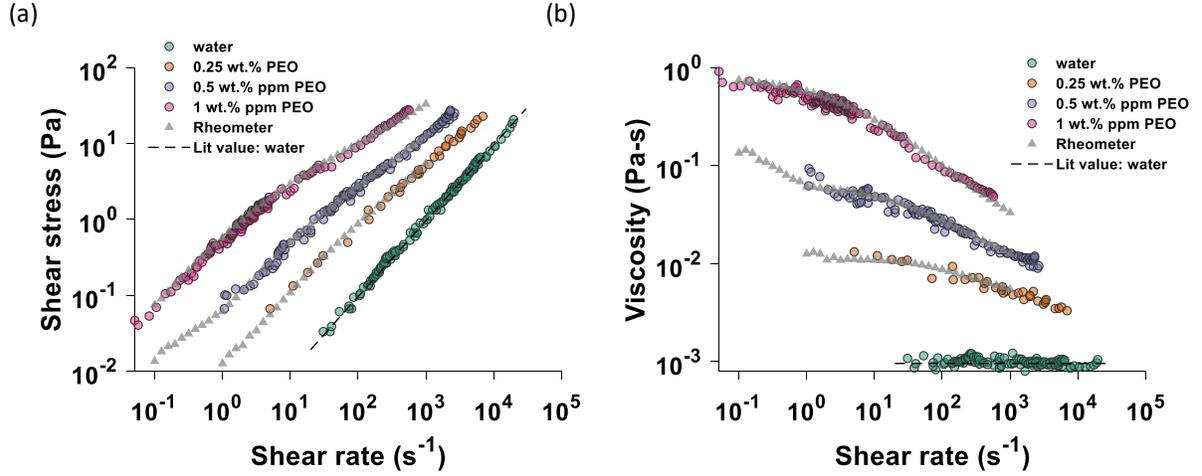

**Figure 7.** Holographic shear rheology of water and viscoelastic PEO solutions. (a) Shear stress versus shear rate data obtained from DHM-PTV compared with rheometry data (triangles) for PEO solutions and with literature values for water (dashed line). (b) Viscosity versus shear rate data for PEO solutions and water.

### C. Holographic shear rheology from flow kinematics

To quantify shear rheology of viscoelastic fluids, we conducted experiments in the thin-slit geometry and imposed a constant pressure drop. The DHM-PTV analysis pipeline was used to obtain the depth-wise velocity profile and the shear stress and shear rate were calculated as previously discussed in Sec. II A. The shear stress versus shear rate data are shown in Fig. 7a for water as well as 0.25, 0.5 and 1 wt %. aqueous PEO solutions. This data set was obtained from the mid-plane depth-wise velocity profile. Considering depth-wise velocity profiles from other vertical planes within 100 μm of the symmetry plane yielded imperceptible differences. In Fig. 7b, the corresponding shear viscosity versus shear rate data are shown. The measured data is in good agreement with that obtained from standard rheometry for viscoelastic fluids, and literature values for water. Thus, our HSR approach is well-suited for characterizing the shear rheology of viscoelastic fluids.

To obtain the data shown in Fig. 7, we typically used three different inlet pressure conditions, with a single pressure condition yielding an order of magnitude variation in shear rate. We find that the shear rates ranged from ~ 0.05 s$^{-1}$ to 20,000 s$^{-1}$, with the lowest shear rates accessible only for the high-viscosity fluids. With high-viscosity fluids, the depth-wise velocity variation is gradual compared to the low-viscosity fluids enabling access to lower shear rates. It is interesting to note that in microfluidic viscometry approaches reported to date [16] , the flow rates need to be adjusted to small values to access low shear rates, however in our HSR approach this need is obviated since shear rate is determined from the velocity variation.

We now discuss factors that are important in obtaining reliable HSR data. The shear rates in HSR are estimated by numerical differentiation of the digitally reconstructed data which makes them sensitive to noise in regions where velocity changes steeply. As a result, the HSR approach requires robust characterization of velocity variation over small lengths. The smallest length scale is dictated by the 1 μm inter-plane spacing during reconstruction.  This reconstruction granularity coupled with the marginal velocity changes near the peak of the velocity profile limits access to viscosity data as $\dot{\gamma} \to 0$.

Similarly, stress measurements rely on accurate identification of the location of minimum shear $Z_{\dot{\gamma}=0}$. This can be problematic when there is significant deformation in the PDMS channel due to strong pressure-driven flow [76]. At the highest imposed pressures (~ 355 mbar), we observed a maximum channel deformation of ~ 2 - 3 μm near the center of the thin slit which correlates well with estimates from the analytical expression $h(x) = \frac{3}{2} h_0 \left(1 + \frac{\alpha P(x) w}{E h_0}\right)$ where $h$(x) and $h_0$ denote the maximum deformed height and the undeformed height, $P(x)$ is the local pressure, E is the Young's modulus and $\alpha$ is a proportionality constant [77]. Nevertheless, we observe that the results from HSR are in good concordance with macrorheology, indicating that these small channel deformations do not strongly impact HSR data. This is because we estimated our $Z_{\dot{\gamma}=0}$ from the velocity field data rather than depending on post fabrication characterization of the channel geometry. Finally, we note that the noise in the estimation of shear stress and shear rate is independent of each other and can result in accumulated noise in estimation of viscosity as it depends on the ratio of the two values.

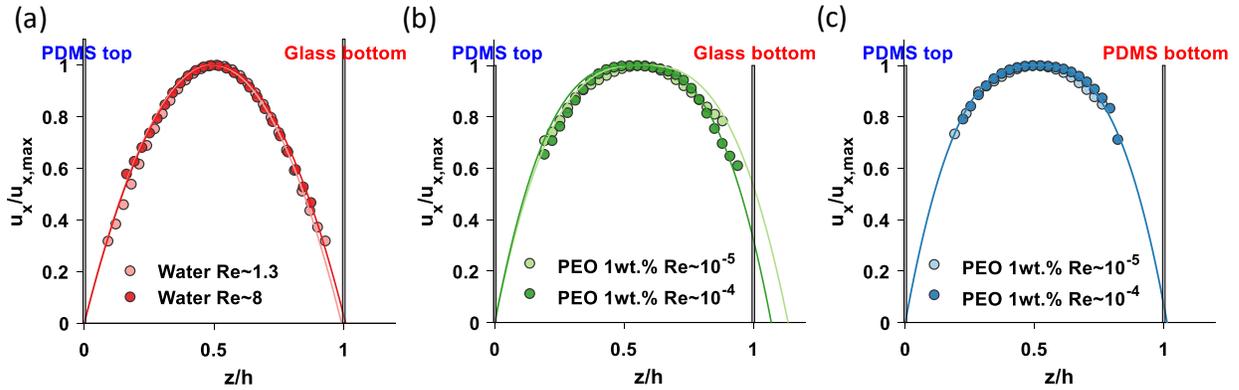

**Figure 8.** Characterization of wall slip in microfluidic flows. Normalized depth-wise velocity profiles at the mid-plane for (a) water in a PDMS channel bonded to glass slide (b) PEO in a PDMS channel bonded to glass slide (c) PEO in a PDMS channel bonded to PDMS coated glass slide. The lines are the fits to Eqn. (5)

### D. Additional considerations for holographic shear rheology

In this section, we discuss two important effects that were apparent when performing holographic shear rheology. The first is associated with slip in viscoelastic flows and the second is particle migration in viscoelastic flows. Both these phenomena were observed under certain experimental conditions. Below, we describe results pertaining to these phenomena, which not only emphasize their importance for HSR but also highlight that DHM-PTV is a powerful tool to analyze these phenomena in viscometric flows.

#### i. Fluidic slip in viscoelastic flows

In standard macrorheometry, wall slip presents a problem for obtaining accurate rheological data and corrections need to be implemented [78]. Since no assumptions about the wall boundary conditions are required in HSR, this method for determining shear rheology is indifferent to the presence of slip. Previously, slip has been explored through variable-gap rheometry [79], surface treatment [80], particle tracking [81], evanescent waves spectroscopy [82] and rheo-NMR [12]. The chemical origins of slip can be attributed to molecular interactions between the fluid and the solid surface such as in polymer melts [83] or superhydrophobic surfaces [84] and its dynamics has been investigated for viscoelastic flows using PTV [85]. Here, we show that slip can occur under certain experimental conditions and this slip can be characterized using DHM-PTV.

In our experiments, the microchannels were made with PDMS replicas bonded to glass slides. In these devices, the velocity profiles were characterized using DHM-PTV. The typical approach for determining slip is to examine if there is truncation in the near-wall velocity profile. As discussed in the next section, in our experiments we observe depletion of seeded particles near the wall due to cross flow migration in viscoelastic flows [75]. Therefore, we are limited to velocity characterization beyond 3 – 5 µm away from the wall. Because of this limitation, we extrapolate near-wall information from the bulk by fitting to a power-law model as:

$$v_z = \left(-\frac{\Delta P}{L_{ch}}\frac{h_{ext}}{K}\right)^{\frac{1}{n}} \frac{h_{ext}}{\left(\frac{1}{n}+1\right)}\left(1 - \left(\left|\frac{z}{h}\right|\right)^{\frac{1}{n}+1}\right) \qquad (5)$$

The depth-wise velocity profile at the midplane is fitted to estimate the span $h_{ext}$ for which the velocity extrapolates to zero. The apparent slip length is characterized as Δh = ($h_{ext}$ – h).

In Fig. 8a, we show the results for flow of water in a thin slit PDMS device bonded to glass at Re = 1.3 and 8. The midplane depth-wise velocity profile extrapolates to zero on the PDMS top surface as well as the glass bottom indicating the absence of slip behavior. However, when 1 wt% PEO solutions was introduced in the same channel at different driving pressures such that Re ≈ $10^{-5}$, $10^{-4}$ and Wi = 1.4, 7.2 respectively, the velocity profiles extrapolated to zero at 3 µm and 5 µm beyond the glass boundary of the channel indicating an apparent slip-like behavior (Fig. 8b). We note that due to viscoelastic particle migration, more velocity data is missing for PEO solution compared to that of water. Previously, Degre et al. [46] reported wall slip in flows of high molecular weight PEO solutions over a glass bottom surface similar to flow systems employed in this study.

Next, we coated the glass substrate with a 50 µm PDMS layer prior to bonding to check if a surface with different chemical interactions with PEO might alter the slip behavior observed with glass surface. In this experiment, the bottom surface was fixed at the same distance from the microscope objective as the uncoated case to ensure a static reference for the velocity fields and 1 wt% solutions of PEO was introduced at the same driving pressures (Re ≈ $10^{-5}$, $10^{-4}$ and Wi = 1.2, 7). The flow of PEO over PDMS coated bottom surface did not exhibit similar slip as in the case of glass indicating the chemical nature of the fluidic slip based on surface properties (Fig. 8c). Thus, our experiments and analysis indicates the presence of a finite slip at small shear rates in case of highly viscoelastic 1 wt. % PEO solution flowing over a glass bottom surface. We also tested PEO solutions of 0.25 and 0.5 wt% in PDMS/glass devices and did not observe slip-like behavior indicating that this phenomenon is more apparent in semi-dilute polymer solutions. Overall, even though the HSR technique is insensitive to the presence of wall-slip, these results indicate that slip may occur in microscale viscoelastic flows, and DHM-PTV is a useful means to observe this phenomenon.

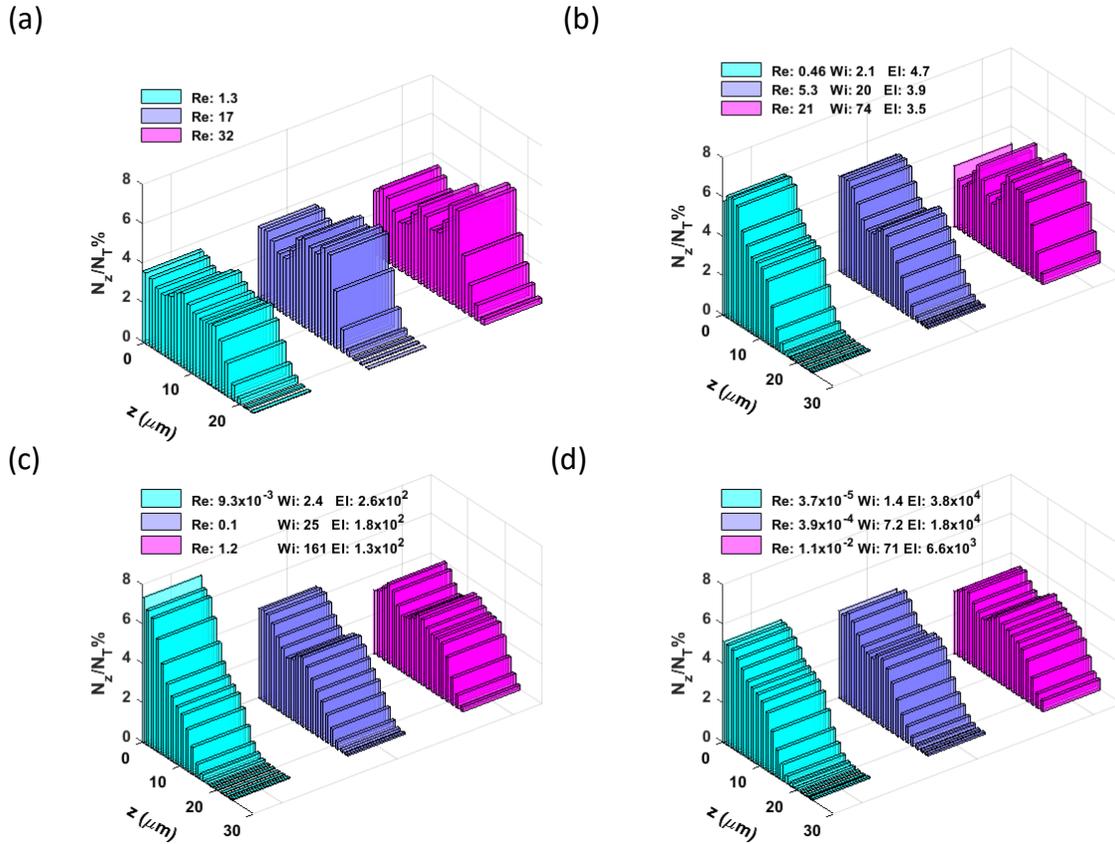

**Figure 9**. Particle distributions are shown in the vertical plane from centerline to top-wall for a thin-slit microchannel. Each fluid is tested at three different pressure-driven flow values (a) Water (b) 0.025 wt% PEO (c) 0.25 wt% PEO and (d) 1 wt% PEO.

## ii.  Particle migration in Newtonian and viscoelastic micro-slit flows

The acquisition of flow kinematics and subsequent rheometry using DHM PTV requires a non-sparse particle field to permeate the flow volume of interest. It is essential, therefore, to recognize any stratification that may occur due to the effects of particle size, flow rate, particle-particle interactions, confinement, slip and rheology of the suspending fluid [75]. As mentioned briefly before we have observed particle stratification in our experiments, which impacted our velocity profile characterization. We discuss these experiments and results below.

Particle migration can occur due to solely inertia [86][ref] as well as due to purely normal stress differences in viscoelastic fluids [71][ref]. In addition, entry junctions promote focusing in Newtonian flows [66, 67]. We investigated whether particle migration can occur under conditions of our rheology studies where both inertial and viscoelastic effects are present. The particle field was analyzed in the thin-slit geometry by counting the number of particles $N_z$ in horizontal planes (of width ~ 50 µm ) and normalizing the counts by the total number of particles $N_T$. Holographic imaging was performed at ~ 115$D_h$ from the entrance which is enough for flow stabilization but significantly lower than typical length scales (> 1000$D_h$) employed for equilibrium particle focusing [75, 87-89]. As a result, the data shown here does not pertain to equilibrium focusing dynamics of particles.

The particle distribution in the vertical midplane from the centerline to the top wall for water is shown in Figure 9a. At low Re (~ 1), we observe a nearly homogenous particle distribution along the vertical midplane whereas at higher Re values (~ 32), the particles are depleted from the core and wall regions and a rise in concentration between the channel center and the walls is detected at approximately ~ 0.25$h$ - 0.3$h$ away from the center. The particle distribution for water at low Re is consistent with a reversible Stokesian flow that does not allow migration while the stratification at higher Re originates from inertial nonlinearities [86].

Next, we consider pressure driven flows of 0.025 wt.% PEO for the viscoelastic case with weak shear thinning as shown in Figure 9b. In the viscoelasticity dominant regime (Re = 0.46, Wi = 2.1, El = 4.7), the particle distribution is pronounced at the center indicating a migration induced by the normal-stress difference. In flow with non-negligible inertia and elasticity (Re = 5.3, Wi = 20, El = 3.9), the peak in the particle distribution at the center is slightly diminished while a secondary maximum appears at ~ 0.25*h* - 0.3*h* indicating the competition between viscoelastic and inertial effects in localized reordering of the particle field. Finally, when Re = 21 and Wi = 71 (El = 3.5) particle depletion occurs at the center with a distinct rise between the center and the wall indicating a more prominent role of inertial effects. We also measured particle distributions in 0.25 wt. % and 1 wt. % PEO solutions which are more viscoelastic and shear thinning that 0.025 wt% PEO solution (Figure 9c, d). However, the trends in particle distribution are very similar to that of 0.025 wt% PEO solution.

In viscoelastic weakly shear thinning flows (0.025 wt. % PEO, Figure 9b) at low Re, the observed migration towards regions of lowest (absolute) shear is similar to a second order fluid [71, 73]. In strongly shear thinning viscoelastic flows (0.25 wt.% and 1 wt.% PEO, Figure 9c-d) the stratification is again similar to a second order fluid at low Re, however at higher Re, the strong shear thinning appears to drive a second maximum between the wall and the center alongside a diminished central peak and redispersion in the distribution profile.

Overall, we find that in holography shear rheology of viscoelastic fluids, particle distribution can be non-uniform which precludes sparse velocity vector field especially close to the walls. Nevertheless, as we have shown HSR is capable of characterizing the shear rheology of viscoelastic fluids.

### E. Conclusions

In summary, digital holography enabled shear rheology of viscoelastic fluids has been demonstrated. An inline holography setup was used and DHM-PTV analysis provided accurate 3D flow kinematics which resulted in measurement of shear viscosity curves. We have shown that this HSR approach can characterize shear rheology of viscoelastic fluids across a wide range of shear rates. From rheometry perspective, HSR obviates the need to have external sensors to measure shear rheology and is not limited by the presence of wall-slip. The holography system presented here can be further miniaturized [90, 91] with the current microfluidic assembly, potentially leading to compact and portable rheometers. Lastly, there is a need to extend the DHM-PTV approach to other complex fluids and geometries. Compared to other velocimetry techniques, the holographic approach does not require mechanical scanning and therefore has significant potential to characterize time-resolved 3D velocity fields, opening up new opportunities in viscoelastic fluid mechanics.

## Acknowledgments

The authors would like to thank Dr. Dhananjay Singh and Dr. Naureen Suteria for help with digital reconstruction and DOS experiments respectively. We are grateful to Dr. Paul Salipante for the useful discussions regarding particle tracking. Finally, we would like to acknowledge Prof. Vinothan Manoharan and his lab for helpful discussions regarding DHM.